\documentclass[acmtog,authorversion,nonacm]{acmart}
\usepackage{ulem}
\usepackage{amsmath}
\usepackage{multirow}
\usepackage{caption}
\usepackage{subcaption}
\usepackage{comment}
\usepackage{diagbox}
\usepackage{algorithm}
\usepackage{algpseudocode}
\usepackage{placeins}
\usepackage{afterpage}
\def\mille{\ensuremath{{}^\text{o}\mkern-5mu/\mkern-3mu_\text{oo}}}

\usepackage{amsthm}
\makeatletter
\def\thm@space@setup{%
  \thm@preskip=20pt
  \thm@postskip=-10pt
}
\makeatother

\theoremstyle{thm}

\newtheorem{definition}{Definition}[section]

\DeclareMathOperator*{\argmin}{arg\,min}

\newcolumntype{P}[1]{>{\centering\arraybackslash}p{#1}}

\AtBeginDocument{%
  \providecommand\BibTeX{{%
    \normalfont B\kern-0.5em{\scshape i\kern-0.25em b}\kern-0.8em\TeX}}}

\setcopyright{none}

\acmSubmissionID{577}

\begin{document}

\title{Computational Design of Wiring Layout on Tight Suits with Minimal Motion Resistance}

\author{Kai Wang}
\affiliation{%
  \institution{Xiamen University}
  \city{Xiamen}
  \state{Fujian}
  \country{China}}
  
\author{Xiaoyu Xu}
\affiliation{%
  \institution{Xiamen University}
  \city{Xiamen}
  \state{Fujian}
  \country{China}}
  
\author{Yinping Zheng}
\affiliation{%
  \institution{Xiamen University}
  \city{Xiamen}
  \state{Fujian}
  \country{China}}
  
\author{Da Zhou}
\affiliation{%
  \institution{Xiamen University}
  \city{Xiamen}
  \state{Fujian}
  \country{China}}
\author{Shihui Guo}
\authornote{Corresponding author: Shihui Guo (guoshihui@xmu.edu.cn)}
\affiliation{%
  \institution{Xiamen University}
  \city{Xiamen}
  \state{Fujian}
  \country{China}}
\author{Yipeng Qin}
\affiliation{%
  \institution{Cardiff University}
  \city{Wales}
  \country{United Kingdom}}
\author{Xiaohu Guo}
\affiliation{%
  \institution{University of Texas at Dallas}
  \city{Dallas}
  \country{United States of America}}

\begin{abstract}
An increasing number of electronics are directly embedded on the clothing to monitor human status (e.g., skeletal motion) or provide haptic feedback.
A specific challenge to prototype and fabricate such a clothing is to design the wiring layout, while minimizing the intervention to human motion.
We address this challenge by formulating the topological optimization problem on the clothing surface as a deformation-weighted Steiner tree problem on a 3D clothing mesh.
Our method proposed an energy function for minimizing strain energy in the wiring area under different motions, regularized by its total length.
We built the physical prototype to verify the effectiveness of our method and conducted user study with participants of both design experts and smart cloth users.
On three types of commercial products of smart clothing, the optimized layout design reduced wire strain energy by an average of 77\% among 248 actions compared to baseline design, and 18\% over the expert design.
\end{abstract}

\ccsdesc[500]{Computing methodologies~Shape analysis}
\ccsdesc[500]{Human-centered computing~Systems and tools for interaction design}

\keywords{Sensor Connection Design, Human Motion Analysis, Steiner Tree Problem}

\begin{teaserfigure}
  \includegraphics[width=\textwidth]{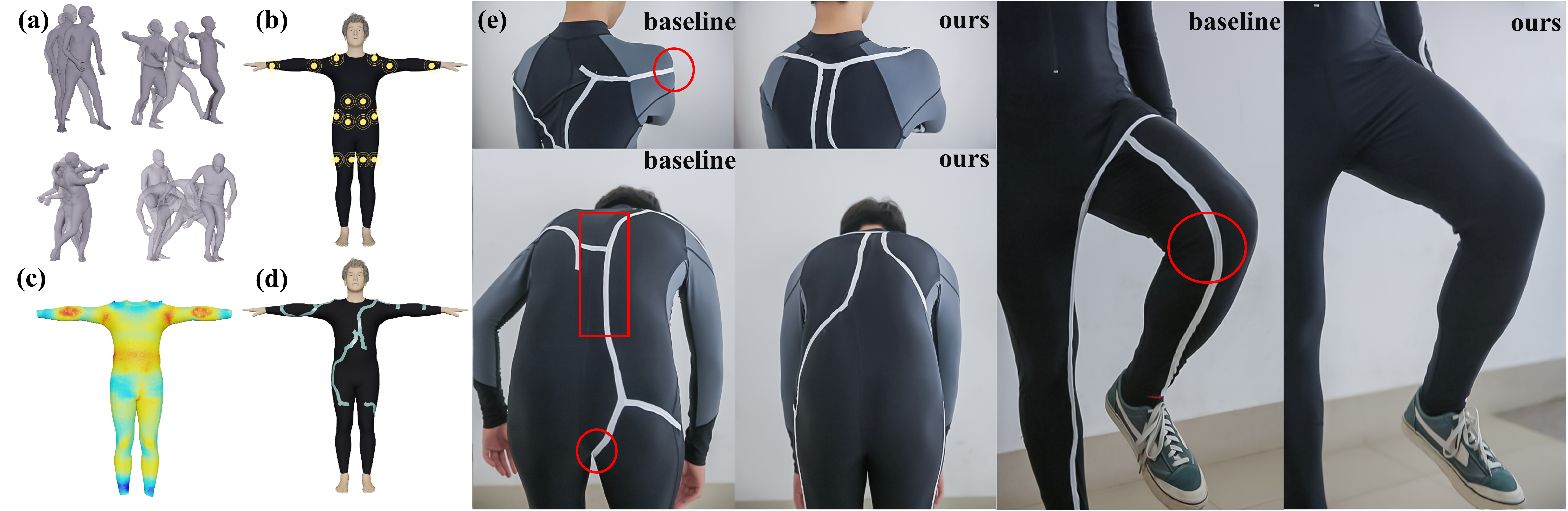}
  \caption{Designing wiring layout on tight suits. 
  Given a set of motion data (a) and on-garment terminals (b), the goal is to find the wire layout connecting the terminals, while minimizing motion resistance when the user is wearing the tight suit.
  We analyze the cloth strain deformation (c), and obtain the wiring layout (d) by resolving a deformation-weighted Steiner tree problem. User experiments show that our approach can effectively reduce the deformation strain at various body parts (highlighted in red boxes) compared with a minimum-length baseline approach (e).}
  \label{fig:teaser}
\end{teaserfigure}

\maketitle

\section{Introduction}
Smart clothes augment traditional garments, by installing a variety of electronic units, including but not limited to IMUs \cite{7319695}, flexible sensors \cite{pointner2020knitted}, pressure sensors \cite{lee2015conductive}, haptic actuators \cite{ramachandran2021smart,lopes2017interactive}, etc. 
Notably, most smart clothes (for example, the commercial products Tesla Suit, Athos) are tight-fitting to guarantee the close conformability with human skin, to avoid sensor displacement and to accurately monitor human status.
Electronic components are connected using either wired or wireless (Bluetooth, WiFi, etc.) approaches \cite{stanley2022review}. 
Wireless connection enjoys its advantage in being cable-free. Compared to wireless connection, the necessity of wired connection for smart clothes stems from its lower susceptibility to environmental interference and easy time synchronization among multiple electronic devices. Currently, the popular adoption of wired connections is evidenced by both previous literature \cite{ancans2021wearable} and in various commercial products like Tesla Suit \cite{tesla}.
Typically, Bluetooth (as one of the most popular wireless solutions) only supports a maximum number of 7 connected devices, which falls short of the demand for smart garment design. 
Therefore, exploring the layout of wiring design offers the benefit for future development of smart clothes.
However, wiring inevitably impacts human motion while motion-driven deformation may also damage wiring connection when being stretched.
Therefore, wire layout greatly affects the mechanical durability, activity performance and user experience when the user is wearing smart clothing \cite{ruckdashel2022smart,agcayazi2018flexible}.
Therefore, it is crucial to optimize on-garment wiring to minimize intervention with user's physical activity, thereby improving user comfort \cite{yokus2016printed}.
Previous works sidestepped this issue by adding redundancies (e.g., ``waviness'', ``interdigitation'') to circuit design.
For example, \citeauthor{vervust2012integration} [2012] developed a novel electronic circuit technology that adds ``waviness'' to the circuits to make them stretchable.
However, these strategies did not directly solve the problem and are suboptimal due to the redundancies they introduce, which require additional manufacturing processes and costs.
In addition, they contradict a common strategy in the fabrication of existing products that encloses wires and straps inside the fabric by hot pressing, whose benefits are three-fold: i) guaranteed structural firmness of the enclosed wire; ii) insulation from skin conductance; iii) waterproof, including the prevention of sweating.
Hence, a direct and automated wiring solution that can improve user comfort without introducing redundancies is desired, but its existence remains an open question.

In general, there are two primary obstacles that hinder the development of a feasible automated solution. Firstly, the mathematical model required for the solution has not yet been constructed. Secondly, there is a lack of a reliable method for quantifying the impact caused by motion.
For the first obstacle, a \normalem{minimum spanning tree} constructed on a graph with on-body electronics as vertices and pair-wise shortest paths ({\it i.e.}, wires) as edges is a sub-optimal solution as it does not allow for wire branching, which provides a larger search space and better solutions.
It is also worth noting that a recent work \cite{vechev2022computational} proposed a method to optimize garment design for kinesthetic feedback obtained from electrostatic (ES) clutches. 
While their approach is effective, it is not suitable for our specific task as we solve the problem of wiring layout with a fixed strap width and least motion resistance, rather than determining arbitrarily shaped regions, leading to a completely different solution.
Therefore, effective wiring on garments with respect to a set of on-body electronic components remains an open problem.

In this work, we propose a novel automated wiring method for smart clothing that effectively tackles the aforementioned challenges, including i) a novel formulation of on-garment wiring as a deformation-weighted Steiner tree problem (STP), providing a fresh perspective on the problem; ii) a novel edge weighting scheme that accurately captures the wire deformation resulting from a range of input motions.
We chose STP over other related combinatorial optimization methods like the traveling salesman problem (TSP) and minimum spanning tree (MST) as: STP distinguishes terminal vertices from other vertices on the mesh and only connects them, while TSP and MST treat all vertices equally and traverse them all. Thus, TSP and MST cannot be applied directly, but require a preparatory step to create another topology on top of the mesh, e.g., a graph connecting all pairs of terminal vertices with the shortest paths. However, this is a sub-optimal solution compared to STP because it does not allow branching (a popular example is that it is shorter to connect the three endpoints of an equilateral triangle from its center than to use its two sides).
In our approach, users can specify the positions of electronic components, and the algorithm generates a wire layout to connect the electronic components automatically. 
Our main contributions can be summarized as the following: 
\begin{itemize}
    \item We directly address one of the key challenges in smart clothing design, {\it i.e.}, the automatic wiring problem, by transforming it into the construction of a deformation-weighted Steiner tree (a topology optimization problem) to minimize motion resistance and improve system robustness.
    \item We propose a novel deformation-inspired edge weighting scheme that models the effects of human motion on wire stretching by its area strain and regularizes it with wire length.
    \item In addition to positive subjective feedback, real-world experimental results show that our approach produces wiring layouts that outperform expert designs, reducing wire strain energy by an average of 77\% among 248 actions, which demonstrates the effectiveness of our method.    
\end{itemize}

\begin{figure*}[!h]
  \includegraphics[width=1\linewidth]{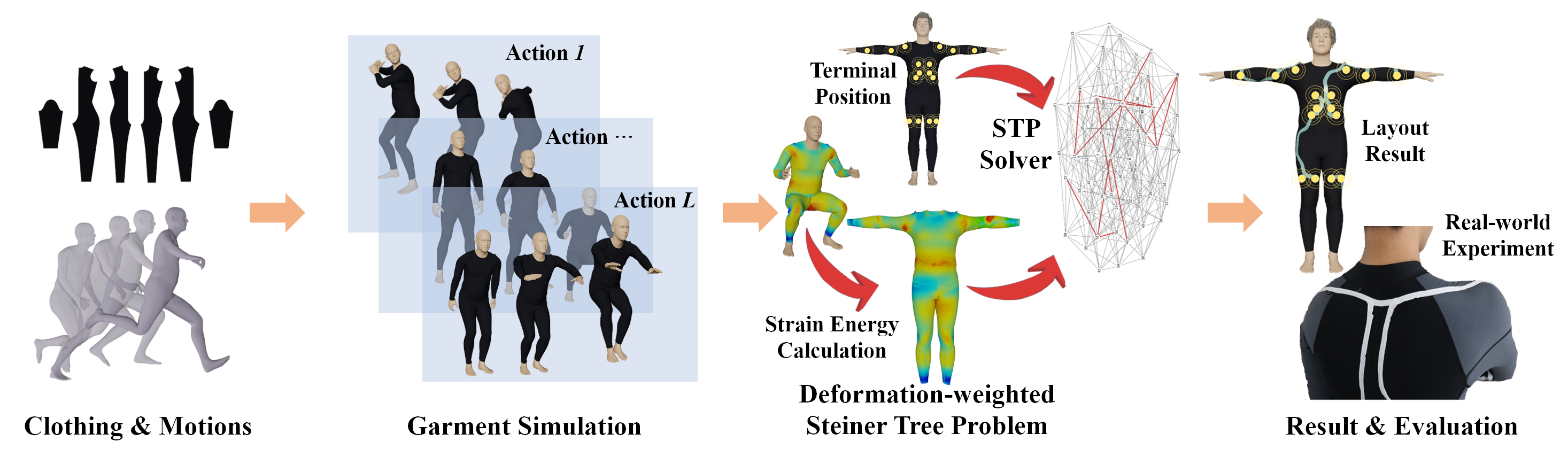}
  \caption{Design pipeline for wiring layout. Starting from the motion sequence and suit pattern model, we first calculate the deformation energy and resolve the deformation-weighted Steiner tree problem, and finally obtained the wiring layout.}
\end{figure*}

\section{Related Work}
\label{sec:related_work}
\paragraph{Design and Application of Smart Clothing}
Smart clothing, as an emerging domain, integrates electronics with textiles to create functional solutions for physiological state monitor \cite{angelucci2021smart}, skeletal motion capture \cite{chen2022full} and body shape reconstruction \cite{zhang2020sensock,chen20213d}. 
Its applications have been demonstrated in fields including rehabilitation \cite{ramachandran2021smart}, assistive walking \cite{kim2019reducing,lee2018autonomous} and interactive entertainment \cite{al2017frozen}.
Its associated physical prototypes include full-body suit \cite{chen2022full}, upper-body shirt \cite{chen20213d,rognon2018flyjacket}, jacket \cite{delazio2018force,gunther2019pneumact}, pants \cite{gholami2019lower}, sock \cite{zhang2020sensock}, elbow pad \cite{liu2019reconstructing,chen2023dispad}, gloves \cite{glauser2019interactive,cai2019demonstration}, etc.

Recent years have seen a booming in commercial products for smart clothing, including Athos, Tesla Suit, Bhaptics, Nadi X, Sensoria Fitness Socks, Siren Diabetic Socks, Ambiotex, AIO smart sleeve, Hexoskin (please refer to the supplementary materials for detailed information).
These products encompass a wide range of monitoring capabilities, covering diverse aspects such as environmental factors, human activity metrics and key biometrics. 
It can be envisioned that a growing number of innovative technology, and most likely electronic gadgets, will be integrated to enhance the functionality of smart clothing.

Despite the increasing demand for smart clothing, methods to aid the design of smart clothing are relatively under-explored.
Existing software solutions in the clothing industry often overlook the complex integration of textile and electronics.
Our work focuses on addressing
one of these challenges, namely the automatic wiring problem, aiming to minimize interference with human motion.

\paragraph{Deformation Analysis for Garment Design} 
Garment deformation caused by human motion is critical for both virtual fitting~\cite{pan2022predicting,hu2022mmtrans,shi2021automatic,wu2022gpu} and real-world clothing design~\cite{sati2021digisew,wolff2022designing,kwok2015styling}.
Therefore, existing studies have incorporated the impact of body movement on fabric simulation into clothing design to simulate the potential impact in real-world situations ~\cite{sati2021digisew,deng2016cloth,li2011customizing}. 
Wolff et al. propose an interactive design tool for creating custom clothing based on 3D body scanning of the intended wearer, by optimizing the shape of the clothing through interpolation and movement of different input postures \cite{wolff2022designing}. 
Liu et al. consider comfort during physical activity and optimize knitted clothing through careful control of the distribution of elasticity to reduce uncomfortable pressure and unnecessary slippage caused by physical activity \cite{liu2021knitting}. 
Pietroni et al. propose an optimization-driven method for automating, based on physics, pattern design for tight clothing \cite{pietroni2022computational}. 
These recent works offer in-depth investigation on the intertwined relationship between human posture and garment design, particularly tight suits.
Another recent work extends to the design of smart clothing, and optimizes the connection structure of electrostatic (ES) clutches for clothing design that provides haptic feedback \cite{vechev2022computational}. 
\citeauthor{montes2020computational} [2020] proposed an optimization-driven approach for automated, physics-based
pattern design for tight-fitting clothing. However, they focused on the stretching of the clothing rather than the wires.
Our work addresses the distinct problem from existing works, i.e., wire layout on tight suits.

We systematically analyzed the garment deformation, in the focused case of tight suit, under more than 248 motion sequences, and aimed to minimize the motion resistance on the deformation surface.

\paragraph{Steiner Tree Problem (STP)}
The definition of STP extends the shortest path problem and minimum spanning tree problem, and has been widely applied in fields such as path planning and circuit design. 
We formulate our problem as an STP variant, i.e., the electronic components as terminals and the wiring layout as paths.
Unfortunately, there is still no polynomial-time algorithm for the general case of the Steiner tree problem \cite{ljubic2021solving}, so the solution algorithms for the problem mainly consist of precise algorithms \cite{dreyfus1971steiner,nederlof2009fast}  and approximation algorithms \cite{gropl2001approximation,robins2005tighter}.
The precise algorithms mainly rely on dynamic programming (DP) to list and compare all possible solutions. 
Their time complexity is mainly determined by the number of terminals $|R|$, the number of vertices $|V|$, and the number of edges $|E|$, and the result can be obtained in exponential time. 
The Dreyfus-Wagner algorithm \cite{dreyfus1971steiner,levin1971algorithm} is a classical precise algorithm based on dynamic programming, with a running time of $O(3^{|R|}·|V|)$. 
Subsequently, \citeauthor{fuchs2007dynamic} [2007] and \citeauthor{nederlof2009fast} [2009] provide improvements on this algorithm. 
The latter two methods calculate the worst-case time algorithm complexity for the special case where all edge weights come from the set ${1, 2, ..., c_{max}}$, and they obtain an upper bound of $O(2·|R|·|V|^2·c_{max} + |V|·|E|·c_{max})$ for the running time. 
Another DP method produces the running time bounds for a wide range of $|R|/|V|$ values \cite{vygen2011faster}. 
Therefore, in cases where the number of terminals is limited and relatively small, precise algorithms can achieve extremely high accuracy. 
In addition, DP algorithms for tree decomposition problems also use a polynomial time algorithm for $|V|$ and an exponential polynomial time algorithm for the tree width, resulting in a polynomial time algorithm for graphs with limited tree width \cite{bodlaender2015deterministic,chimani2012improved,hougardy2017dijkstra}. 
We adopt a precise STP solver \cite{iwata2019separator}, which made dedicated efforts for the case of few terminal nodes, achieving extremely high computational efficiency.

Our work solves the topology optimization problem on the garment deformation surface, by formulating it as a Steiner tree problem.
We discretize it into the clothing mesh and propose a novel energy to comprehensively consider both wire elongation and area strain.

\section{Unobtrusive Wiring on Garments}
\label{sec:layout}

Our goal is to create a novel solution for unobtrusive wiring layout on tight suits, which houses multiple electronics. 
The primary aim of our solution is to minimize motion resistance during physical activities, ensuring an unobtrusive user experience.

\subsection{Problem Definition}

\begin{definition}[Unobtrusive Wiring on Smooth Surfaces]
Given a smooth surface $S \in \mathbb{R}^3$ with $N_p$ terminals $P=\{p_1, p_2, ..., p_{N_p}\}$, unobtrusive wiring is achieved by constructing a Steiner Tree $T^*\in S$ whose strain energy (edge weights) $\mathcal{E}_{T}$ is minimized according to a set of motions $M=\{m_1, m_2, ..., m_L\}$ with length $L$:
\begin{equation}
\begin{aligned}
    T^* = \argmin_{T} \mathcal{E}_{T}(m_1, m_2, ..., m_L).
\end{aligned}
\end{equation}
\label{def:smooth_surface_wiring}
\end{definition}
Strain energy is the elastic potential energy gained by a wire during elongation with a tensile (stretching).

Given the intricate nature of the surface $S$ and motion set $M$ in real-world scenarios, although Definition \ref{def:smooth_surface_wiring} is precise, it must be discretized for effective numerical analysis. To avoid confusion, we use similar notations for the corresponding concepts in both definitions as follows.
\begin{definition}[Unobtrusive Wiring on Discrete Surfaces]
Given a polygon mesh $S = (V,E) \in \mathbb{R}^3$ where $V$ and $E$ are the sets of vertices and edges of $S$ respectively, $N_p$ terminals $P=\{p_1, p_2, ..., p_{N_p}\} \subset V$, unobtrusive wiring is achieved by constructing a Steiner tree $T^*=(P, E')$ with $E' \subset E$ whose strain energy (edge weights) $\mathcal{E}_{T}$ is minimized according to a set of motions $M=\{m_1, m_2, ..., m_L\}$ with length $L$:
\begin{equation}
\begin{aligned}
    T^* = \argmin_{T} \mathcal{E}_{T}(m_1, m_2, ..., m_L).
\end{aligned}
\end{equation}
\label{def:discrete_surface_wiring}
\end{definition}

It is worth noting that in Definition~\ref{def:discrete_surface_wiring}, rather than solely focusing on the surface, we simultaneously discretize the underlying surface into polygon meshes and the wiring Steiner tree into its graph-based version.
This approach stems from the fact that, to the best of our knowledge, there are currently no efficient algorithms available to solve this particular problem, namely the Steiner Tree Problem on arbitrary polygon surfaces \cite{caffarelli2014steiner}.
Thus, we believe that our definition represents the most suitable approach at this juncture. 
Nevertheless, we maintain an optimistic outlook for future breakthroughs that may inspire new variations of Definition~\ref{def:discrete_surface_wiring}.

\vspace{2mm}
\noindent \textbf{Remark.} Definition~\ref{def:discrete_surface_wiring} is also applicable to cases where $p_i$ is placed within the faces of $S$. In such cases, we can subdivide the corresponding face, add $p_i$ as a new vertex in $V$, and include the newly generated edges in $E$.

\subsection{Problem Input}
The input includes:
\begin{enumerate}
    \item A set of motion sequences: $M=\{m_1, m_2, ..., m_L\}$, where a motion sequence $m_i$ is a set of continuous frames of human posture: $m_i=\{ m_{i}^1,m_{i}^2,...,m_{i}^k\}$.
    \item A tight suit model: it includes a 3D mesh model as $S$ and a 2D pattern model. In our current work we use three different sizes of the suit: M, L, XL.
    \item A resting state: we use the 2D pattern model as the resting state to measure the deformation, since the 2D pattern mesh is flat and does not involve any deformation.
    \item A set of terminals $P=\{p_1, p_2, ..., p_{N_p}\}$, indicating the electronics to be connected.
\end{enumerate}

\subsection{Deformation-weighted Steiner Tree Problem}
\label{sec:Motion-weighted Steiner Tree Problem}

\paragraph{Cloth-based Deformation Simulation}
To obtain the surface deformation when performing various movements, we conduct cloth simulation on a tight suit mesh $S$, which is worn by a virtual character when performing the motion in the motion set $M$. 
We import the motion sequence generated by the standard human body SMPL \cite{pavlakos2019expressive} into Marvelous Designer \cite{marverlous} for cloth simulation.
Three SMPL models are generated by matching the body girths on SMPL models \cite{chen20213d} according to the expected values of standard garment size.
For utmost realism, we used the cloth simulation engine from the mainstream cloth design software of Marvelous Designer, which captures friction/displacement between a tight suit and human skin. Detailed simulation parameters can be found in Supplementary Sec.~7. To validate the effectiveness of our simulation, we use the strain sensing system to measure strain deformation on real cloth, which reveals consistency with the simulation.

\subsubsection{Deformation-inspired Edge Weights} 
As specified in Definition~\ref{def:discrete_surface_wiring}, the key challenge to maintaining an unobtrusive user experience during physical activities is to minimize the strain energy $\mathcal{E}$ exerted by motions $M$ onto the on-garment wires $T^*$.
Since $T^*$ is essentially a Steiner Tree in mesh $S$, $\mathcal{E}$ can be decomposed into the individual contribution of each edge $e \in E' \subset E$.
Hence, we focus on calculating the strain energy $\omega(e)$ for each edge $e$ of mesh $S$, which will serve as the edge weights during the construction of Steiner Trees:
\begin{equation}
    \omega(e)=\sum_{f \cap \delta(e) \neq \emptyset} \left( \epsilon\left(f,M\right)+\eta\right)*area\left(f \cap \delta\left(e\right)\right),
\label{edgeweight}
\end{equation}
where $f$ denotes the faces of mesh $S$, $\eta$ is a regularization term of wire length to avoid excessive redundancy of the wire. 
A smaller value of $\epsilon\left(f,M\right)$ selects straighter/shorter wires. 
$area$ denotes the area of the selected mesh, i.e., the area of the cloth straps accommodating wires that are fixed to smart clothes, which should be avoided from stretching. This is important as wires are secured to the cloth straps, thus stretching the cloth straps both along and perpendicular to the direction of the wires will affect its performance.
$\delta(e)$ denotes a local neighborhood of $e$ ({\it e.g.}, of a rectangular shape with $e$ as the midline) which simulates the width of the wire strip in real-world scenarios. 
We compute the intersection $f \cap \delta\left(e\right)$ using the Shapely library~\cite{shapely2007}.
$\epsilon\left(f,M\right)$ is the strain energy of $f$ with respect to motions $M$: 
\begin{equation}
\epsilon(f,M) =\max_{m_i \in M}{ \sum_{j=1}^{k_i} \frac{\epsilon(f, m_{i}^j)}{k_i}}, 
\end{equation}
where $k_i$ is the number of frames in motion sequence $m_i$ and we follow the St. Venant-Kirchhoff model introduced in  \cite{liu2016towards} to calculate $\epsilon(f, m_i)$:
\begin{equation}
\begin{aligned}
    & \epsilon(f, m_i) = \mu\|\mathcal{G}_{f,i}\|_2+\frac{\lambda}{2}tr(\mathcal{G}_{f,i}), \\
    \mu &=\frac{E}{2(1+\nu)}, \lambda=\frac{E\nu}{(1+\nu)(1-2\nu)},
\end{aligned}
\end{equation}
where we set the Young's modulus $E$ of the clothing to 5.4 MPa and the Poisson's ratio $\nu$ to 0.33 \cite{montes2020computational}, and calculate the Green strain tensor $\mathcal{G}_{f,i}$ as:
\begin{equation}
\mathcal{G}_{f,i} = \frac{1}{2}(F_{f,i}^TF_{f,i}-I), 
\end{equation}
where the deformation matrix $F_{f,i}$ is obtained from the difference between $f_{\mathrm{ref}}$ and $f_i$, where $f_{\mathrm{ref}}$ is the shape of $f$ in a {\it reference state} and $f_i$ is the shape of $f$ in motion $m_i$.
$I$ is the identify matrix. 
In Eq.~\ref{edgeweight}, $\omega(e)$ represents the energy sum in the area, and $\epsilon(f,M)$ represents the energy density, which needs to be multiplied by the area in order to get the energy of the wire-attached area (with fixed width).

\subsubsection{Steiner Tree Solver} 
With edge weights $\omega(e)$ defined, we can effectively address unobtrusive wiring on garments using state-of-the-art Steiner Tree algorithms. In this paper, we employ a highly efficient method \cite{iwata2019separator}, which leverages dynamic programming and pruning techniques.

\subsection{Curve Smoothing \& Generation}
The computed Steiner Tree is a set of connected edges on the mesh $S$. 
However, its polyline nature often leads to non-smooth and sharp bending points at endpoints, hindering the wiring in real-world scenarios.
Addressing this issue, we propose to iteratively smooth the computed Steiner tree with spline curves ({\it e.g.,} circular arcs) on the 2D pattern model until its curvature $\kappa$ satisfies:
    $\kappa < \frac{2}{wd}$,
where $wd$ is the width of the wire strip on the garment. 
This ensures that the wires do not self-overlap locally. 
In our experiments, we observed that this requirement can be easily satisfied in a few iterations.
Please see the supplementary materials for details of our iterative smoothing strategy.

\section{Garment Fabrication}
The fabric of the tight suit is made from 82\% nylon and 18\% polyester.
The fabrication is assisted by a tailor with experiences of over 20 years.
The strip is hot stamped onto the fabric, at the temperature of 160 Celsius with the heat-fusion press machine of Maike MK-TH40*60.
The strip width $wd$ is 1.5 cm. 

\section{Experimental Results}
\subsection{Implementation Details}
\paragraph{Hardware \& Software}
We conducted experiments on a standard Windows PC with an Intel(R) Core(TM) i7-12700H (2.30 GHz) CPU.
We implemented our algorithm using Python and developed a visualization interface with Unity3D.
We will release all source code and datasets to the public upon acceptance.

\paragraph{Motion Dataset}

We employed a standard human SMPL model to represent the human body mesh. To facilitate character rigging and animation, we utilized the automatic rigging function provided by the Mixamo website.
Our dataset encompassed a total of 248 human action files, encompassing a diverse range of sports and dances (such as boxing, baseball, jazz dance, etc.). The combined duration of these action files amounted to 40,108 frames, equivalent to 1,336.9 seconds. These action files consist of either periodic actions with multiple cycles of actions or discrete actions with complete performances.
Please see the supplementary materials for more details on the type and duration of motions.

\paragraph{Time Costs.}
The time costs associated with our solution can be divided into three parts:
    i) {\it Calculation of deformation energy.} Given a set of motion sequences $M$, we compute the deformation energy for each face of the human body mesh in every frame. This computation process requires approximately 2.62 seconds per frame.
    Please note that the time cost mentioned is solely dependent on the input motion and is independent of the wiring. As a result, it can be reused for multiple wiring designs.  
    ii) {\it Calculation of edge weights.} As aforementioned, we calculate our deformation-inspired edge weights $\omega(e)$ using Eq.~\ref{edgeweight}. 
    This part takes approximately 24 minutes for a mesh with 23,514 vertexes (46,912 faces and 70,430 edges). 
    This preprocessing step only needs to be computed once per mesh per motion set and can be reused for different wiring layouts (i.e., low amortized cost).
    Same as above, this part is solely dependent on the input motion and can be reused for multiple wiring designs.
    iii) {\it Construction of Steiner tree.} Given the calculated edge weights $\omega(e)$ and user-defined terminals, the Steiner Tree can be constructed in a short time, taking 1.80s for 36 terminals on a mesh with 23,514 vertexes.

\subsection{Evaluation Baseline \& Metrics}
\label{sec:evaluation_baseline}

\paragraph{Baseline.} To demonstrate the effectiveness of our wiring solution, we compare it with a naive baseline, namely {\it minimum length}, which adheres to standard manufacturing requirements and aims to minimize the total wire length without taking motion resistance into consideration.
Specifically, our baseline used a non-deformation-weighted version of the proposed solution (i.e., the original STP solution producing wiring layouts with minimum total lengths) to justify the effectiveness of our deformation weighting scheme (Sec.~\ref{sec:Motion-weighted Steiner Tree Problem}). That is, instead of using edge weights (Eq.~\ref{edgeweight}), the baseline uses $\omega(e)=1$ for all $e$ to compute a minimum-length solution.

\paragraph{Metrics.} Given that our work encompasses both simulation and real-world experiments, we utilize a set of three evaluation metrics. Two of these metrics are employed to assess the simulation results, while the third metric is utilized to evaluate real-world performance:
\begin{itemize}
    \item {\it Maximum Wire Elongation Rate (for simulation).} Given a set of motion sequences, this metric quantifies the maximal amount of elongation experienced by the wire relative to its original length. Please note that we focus solely on the elongation of the wire and disregard any shortening that may occur, as in the context of smart clothing, the wire does not shrink but rather rolls up or deforms along with the clothing.
    \item {\it Deformation Energy (for simulation).} This metric can be viewed as an extension of the edge weights (as defined in Eq.~\ref{edgeweight}) applied to the smoothed curves. In short, we extend the rectangles used in Eq.~\ref{edgeweight} to trapezoidals inscribed within the osculating circles of the spline curves (circular arcs).
    Please see the Supplementary Sec.~5 for more details.
    \item {\it Capacitance Value of Strain Sensor (for real-world scenarios).} 
    To verify the effectiveness of our method in real-world scenarios, we fabricated a strain sensing system, as a prototype of a one-piece tight suit along with 24 capacitive strain sensors (purchased from Elastech).
    The suit size fits a male of 175cm/75kg.
    These sensors were affixed to the tights with the hook-and-loop velcro structure, to capture the clothing deformation during motion. 
    The change in capacitance values of the strain sensors exhibits a highly linear relationship (99.9\%) and high resolution (0.05\%) with respect to strain deformation \cite{yiweiinven2021}. This makes it a reliable indicator of wire elongation in real-world scenarios.
    The sensor reading is wirelessly transmitted to a PC at a frequency of 50Hz, with a digitized value range [0, 1023].
\end{itemize}

\subsection{Body-specific Wiring}

To validate the effectiveness of our method, we examine two scenarios involving a single joint (elbow and knee). We generate two layouts for each scenario: our method and the baseline ({\it minimum length}) specified in Sec.~\ref{sec:evaluation_baseline}.

\begin{figure}[h]
     \centering
     \begin{subfigure}[b]{0.48\linewidth}
         \centering
         \includegraphics[width=\textwidth]{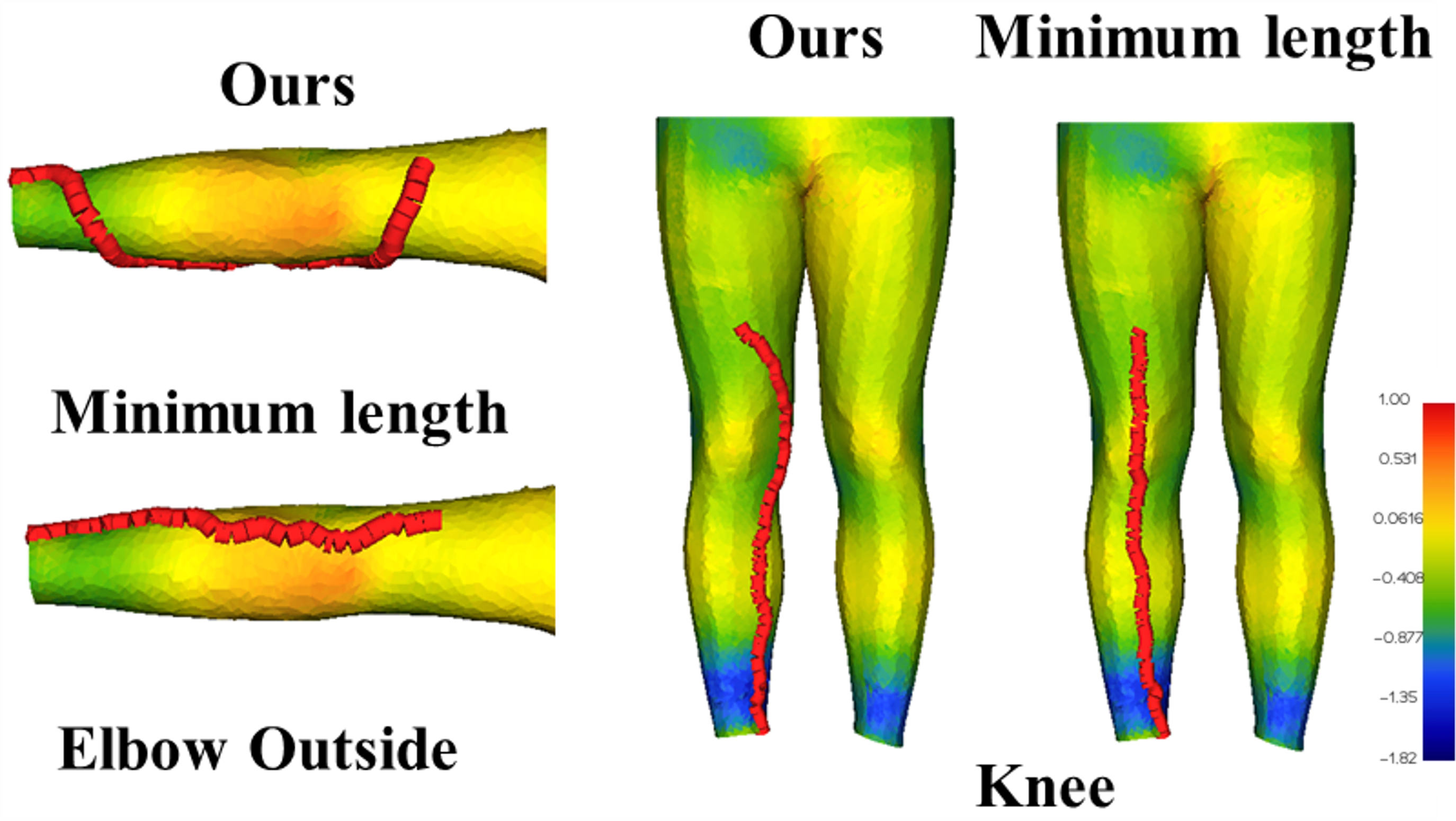}
         \caption{Wire layout}
     \end{subfigure}
     \begin{subfigure}[b]{0.48\linewidth}
         \centering
         \includegraphics[width=\textwidth]{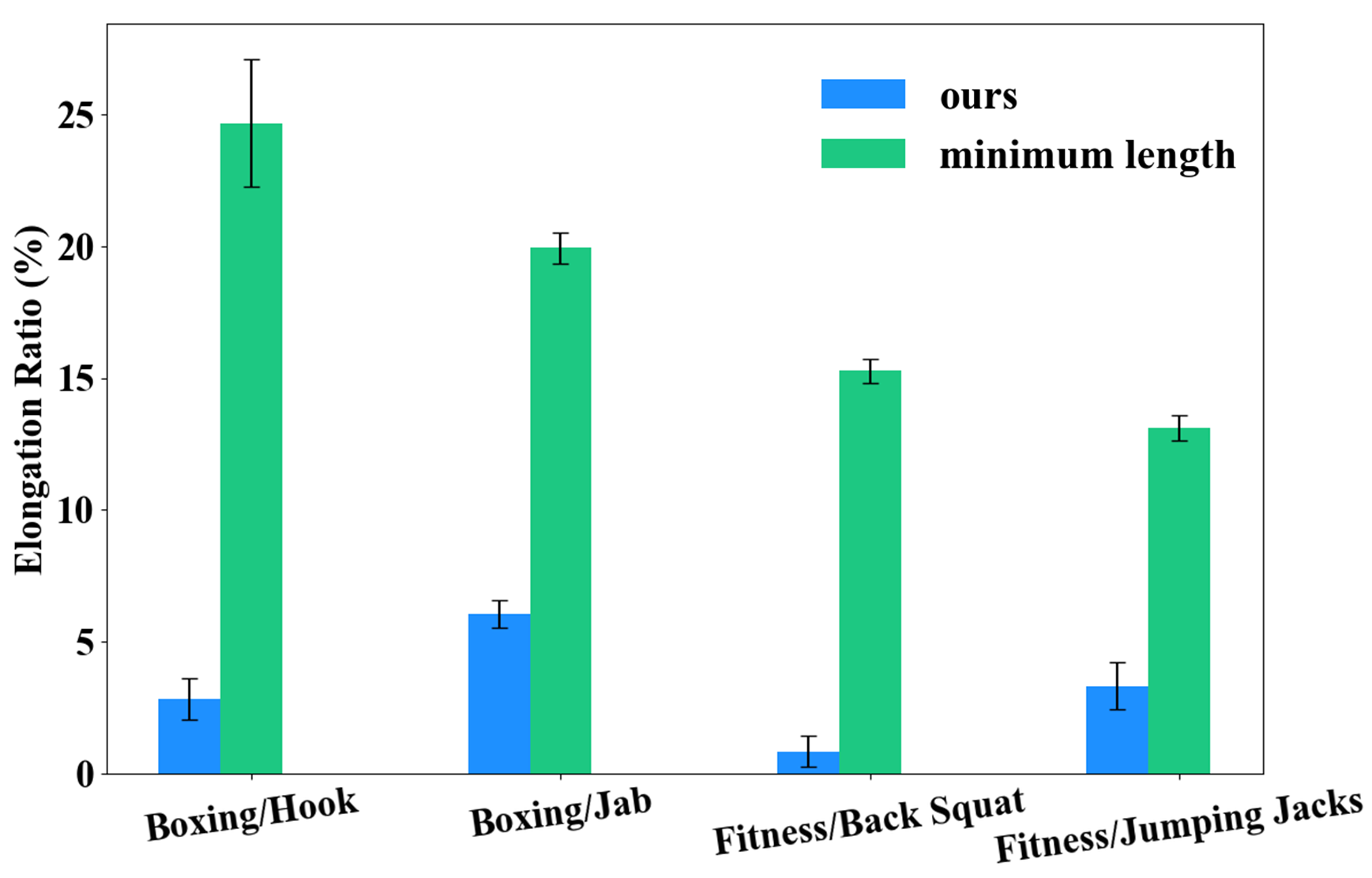}
         \caption{Maximum wire elongation}
     \end{subfigure}
     \caption{Wire layouts (a) and their maximum elongation ratios (b) at the elbow and knee joints, respectively. {The visualization in (a) is based on the deformation energy averaged across multiple elbow/knee joint motions with a logarithmic-scale legend.}}
     \label{fig:single}
\end{figure}

\paragraph{Simulation Experiment}
As Fig.~\ref{fig:single}(a) shows, our solution adeptly avoids regions characterized by high deformation energy, which aligns perfectly with our prior knowledge that the elbow and knee joints exhibit the most significant stretching.
Quantitatively, we compare the maximum wire elongation rates between our method and the minimum length baseline across a variety of elbow-driven ({\it e.g.}, push up) and knee-driven ({\it e.g.}, back squat) motion sequences, respectively. As Fig.~\ref{fig:single}(b) shows, our method consistently outperforms the baseline in all scenarios, showcasing significant improvements.

\begin{figure}[h]
    \centering
    \begin{subfigure}[b]{0.48\linewidth}
         \centering
         \includegraphics[width=\textwidth]{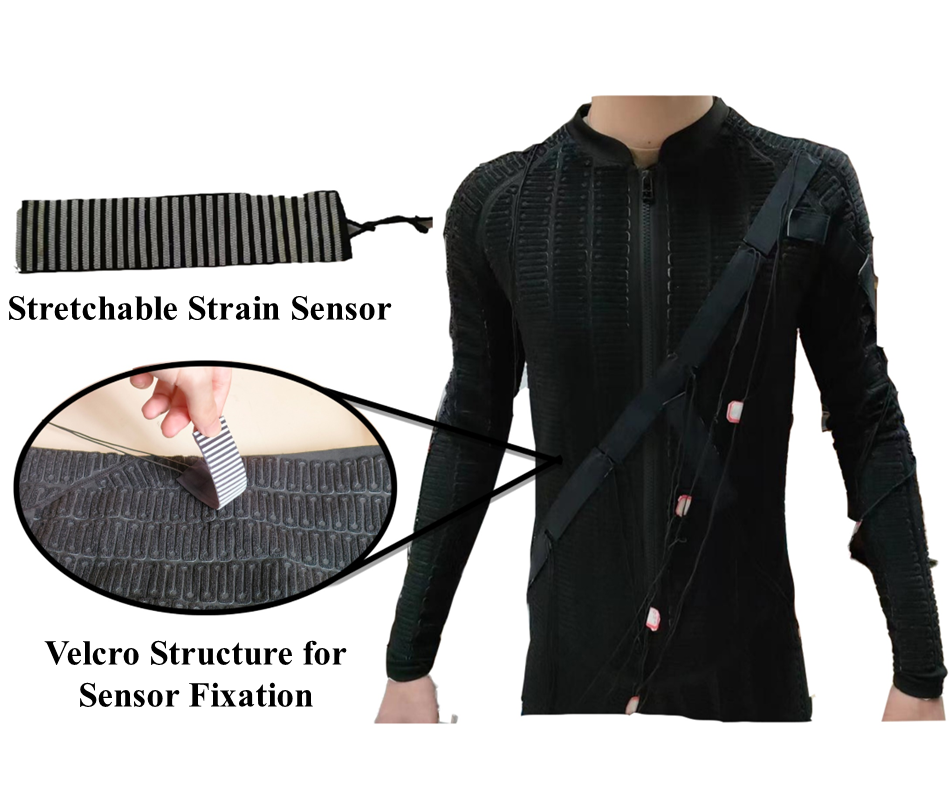}
         \caption{Strain Sensing System}
     \end{subfigure}
     \begin{subfigure}[b]{0.48\linewidth}
         \centering
         \includegraphics[width=\textwidth]{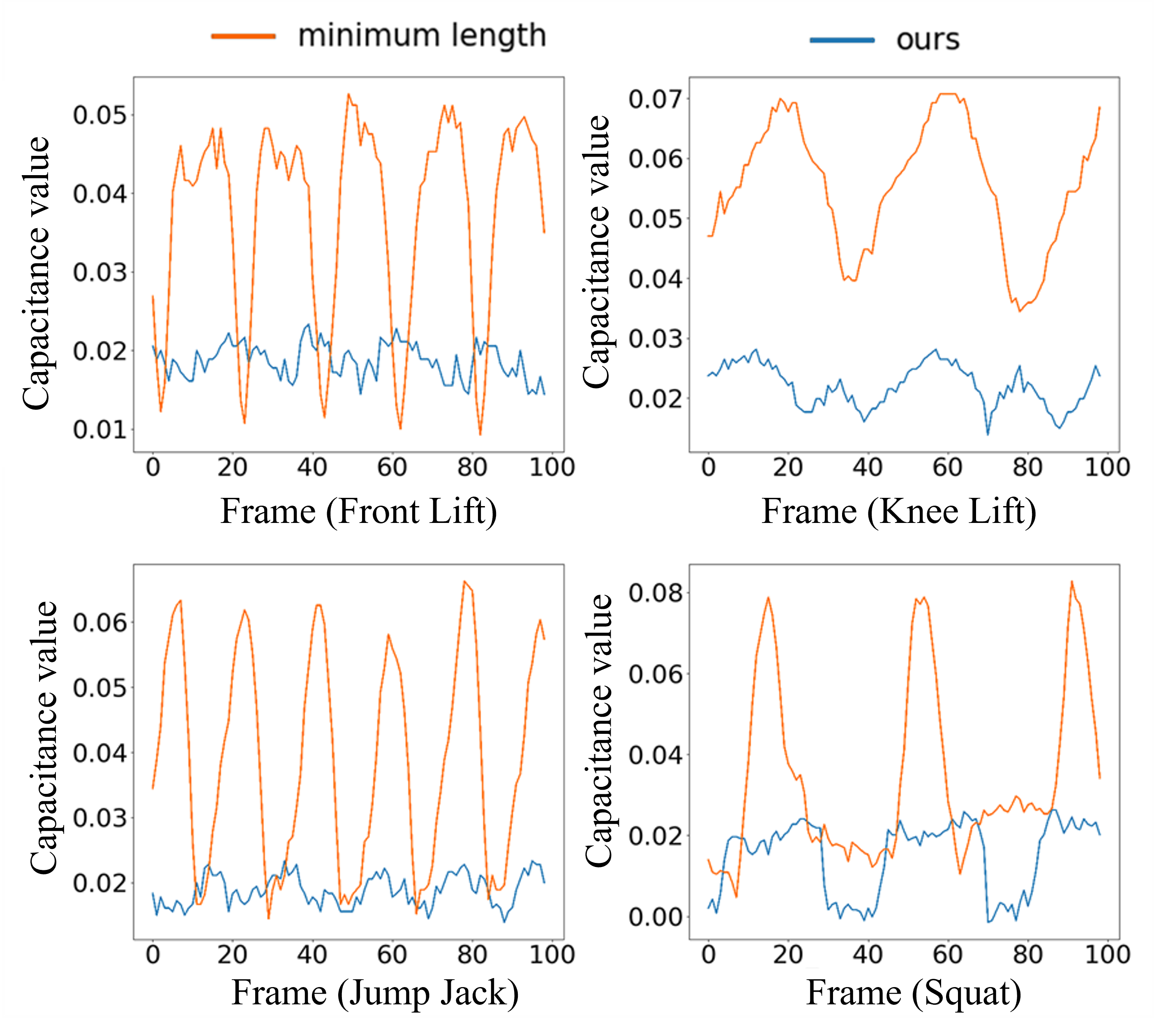}
         \caption{Maximum Strain Deformation}
     \end{subfigure}
    \caption{Real-world experiments to verify the strain deformation. We built a strain sensing system by placing stretchable strain sensor (a) at target layout positions. 
    The sensor capacitance value increases linearly with the magnitude of strain deformation. The results show that our method can effectively reduce strain deformation (b).}
    \label{fig:real_word_4_2}
\end{figure}

\paragraph{Real-world Experiment}
A participant with the suitable body size tried on the strain sensing system (Fig.~\ref{fig:real_word_4_2}(a)), and conducted the aforementioned four motion sequences: hook, push-up, back squat and jumping jacks.
Sensors were placed at the corresponding elbow and knee joints.
As Fig.~\ref{fig:real_word_4_2}(b) shows, compared to the minimum length baseline, our method achieves a significantly lower change in the capacitance values of the strain sensors affixed to the fabricated tights. This observation further substantiates the effectiveness of our wiring solution in minimizing motion resistance.

\subsection{Motion-specific Wiring}

\begin{figure}[h]
  \centering
  \includegraphics[width=0.9\linewidth]{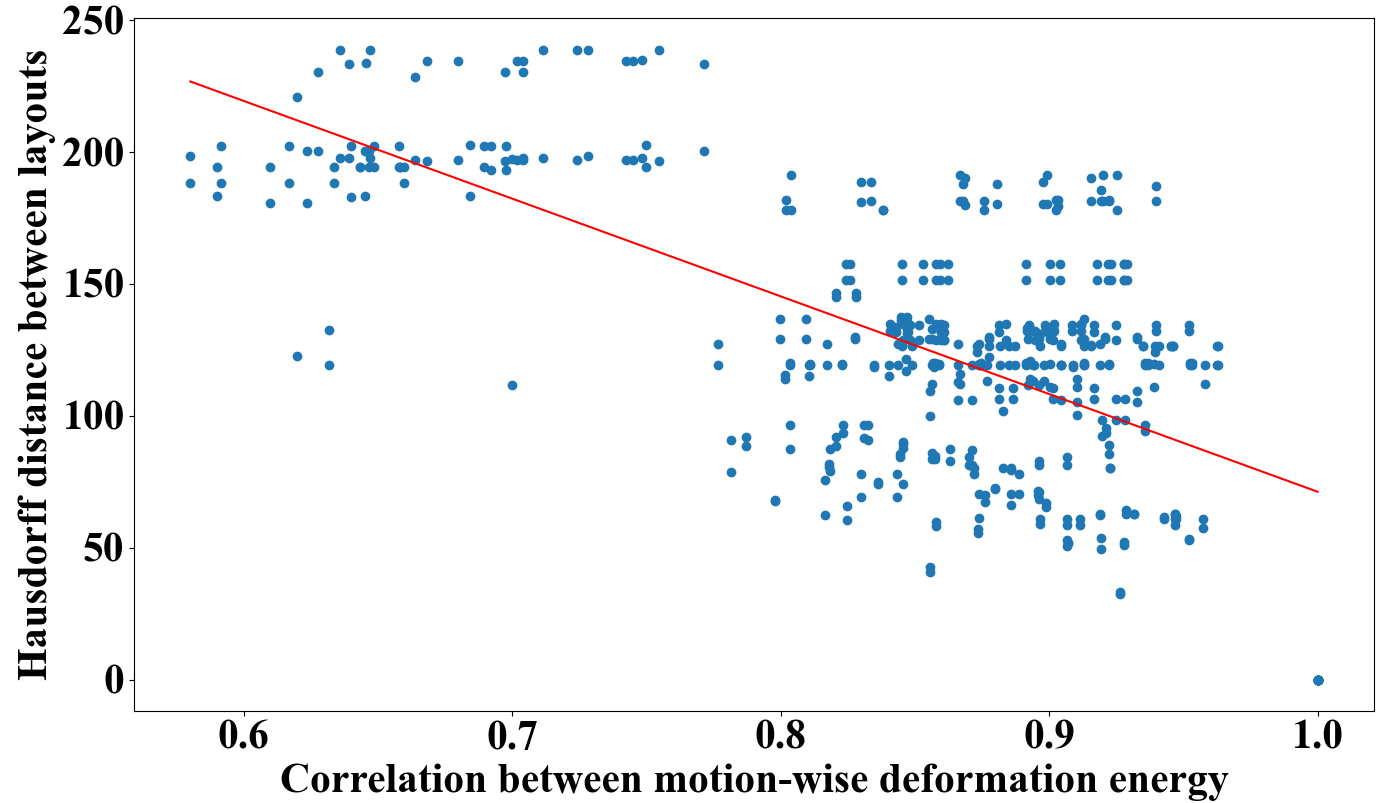}
  \caption{The correlation between motion and wire layout.}
  \label{fig:hausdauff}
\end{figure}

\begin{figure*}[h]
   \includegraphics[width=\linewidth]{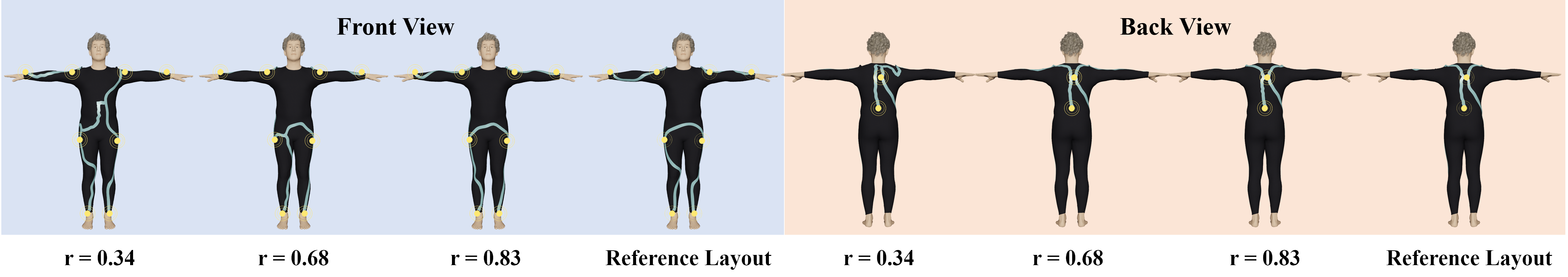}
   \caption{Wiring layout under different levels of motion correlation. The results show that similar motions lead to similar design. 
   Note that the terminal placement is the same across all designs.}
   \label{fig:motion_correlation}
\end{figure*}

Specifically, we explore the correlation between similar motions and the resulting wiring layouts.
We first calculated the deformation energy for each motion sequence, then the Pearson-correlation $R_{M}$ between motion-wise deformation energy.
After obtaining the motion-wise layout, we calculated the Hausdorff distance $D_{L}$ on mesh between wire layouts.
A linear regression between $R_{M}$ and $D_{L}$  (Fig.~\ref{fig:hausdauff}) reveals the correlation coefficient r = -0.66, indicating a negative correlation between the two, i.e., a similar energy distribution leads to a consistent layout.
Fig.~\ref{fig:motion_correlation} visualizes a notable correlation between the similarity of the layouts and the similarity of the weight distribution. 
The aforementioned observations collectively unveil the correlation between motion and wiring layout, showcasing the robustness of our wiring solution.

\subsection{Wiring and Electronics Configurations}

\begin{figure*}
     \includegraphics[width=\textwidth]{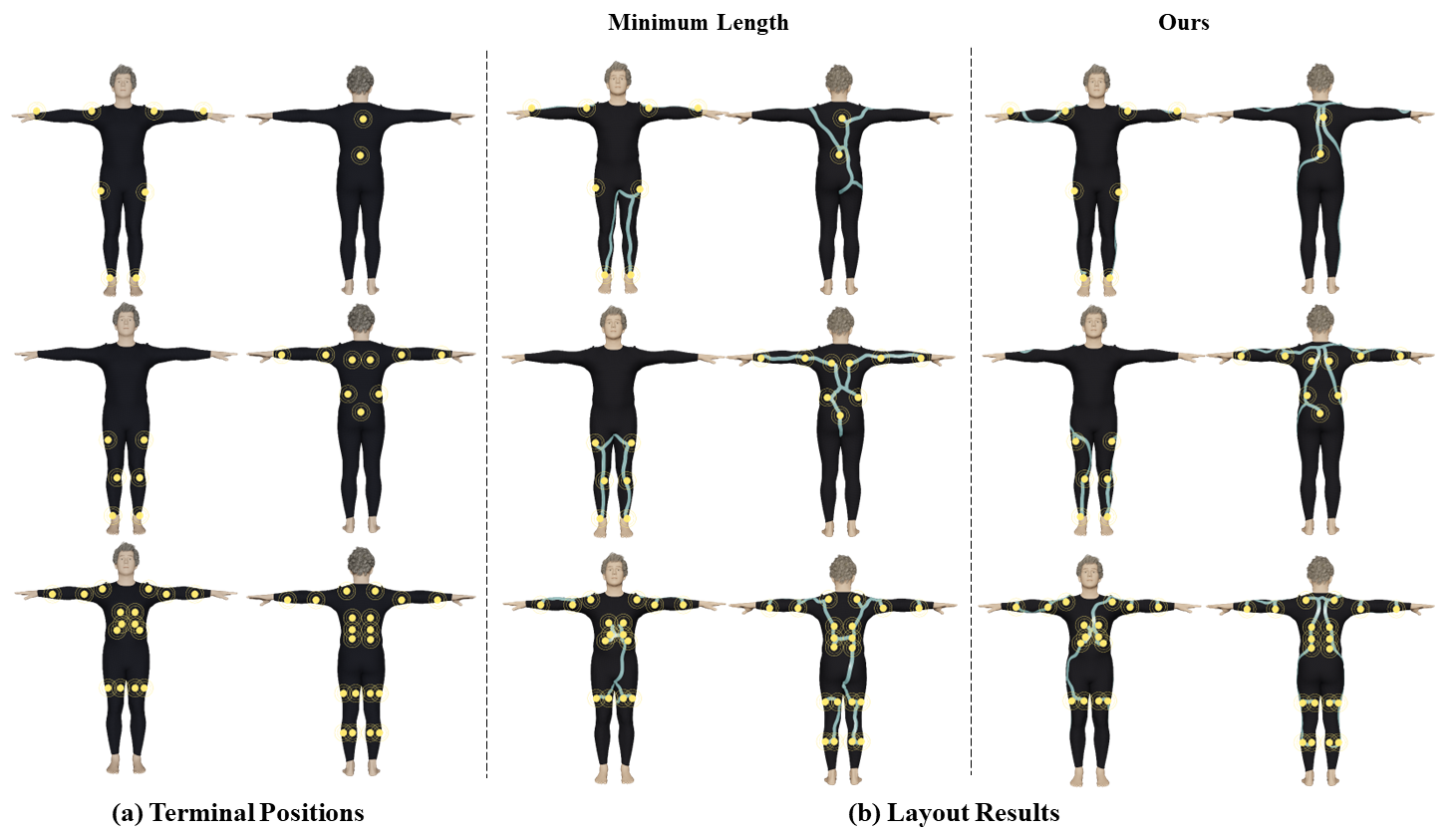}
    \caption{Wiring layout for three terminal configurations under 248 motion sequences.
     Given the terminal positions (a), we obtained the layout results (b).}
     \label{fig:compare_layout}
\end{figure*}

This section evaluates the performance of our wiring solution under three electronics configurations, considering the complete motion collection of 248 sequences.
The placement of electronics was inspired by the corresponding commercial products: Xsens, Tesla Suit Motion Capture and Haptic Feedback (Fig.~\ref{fig:compare_layout}). 

\begin{figure*}[h]
    \centering
    \begin{subfigure}[b]{0.32\linewidth}
         \centering
         \includegraphics[width=\textwidth]{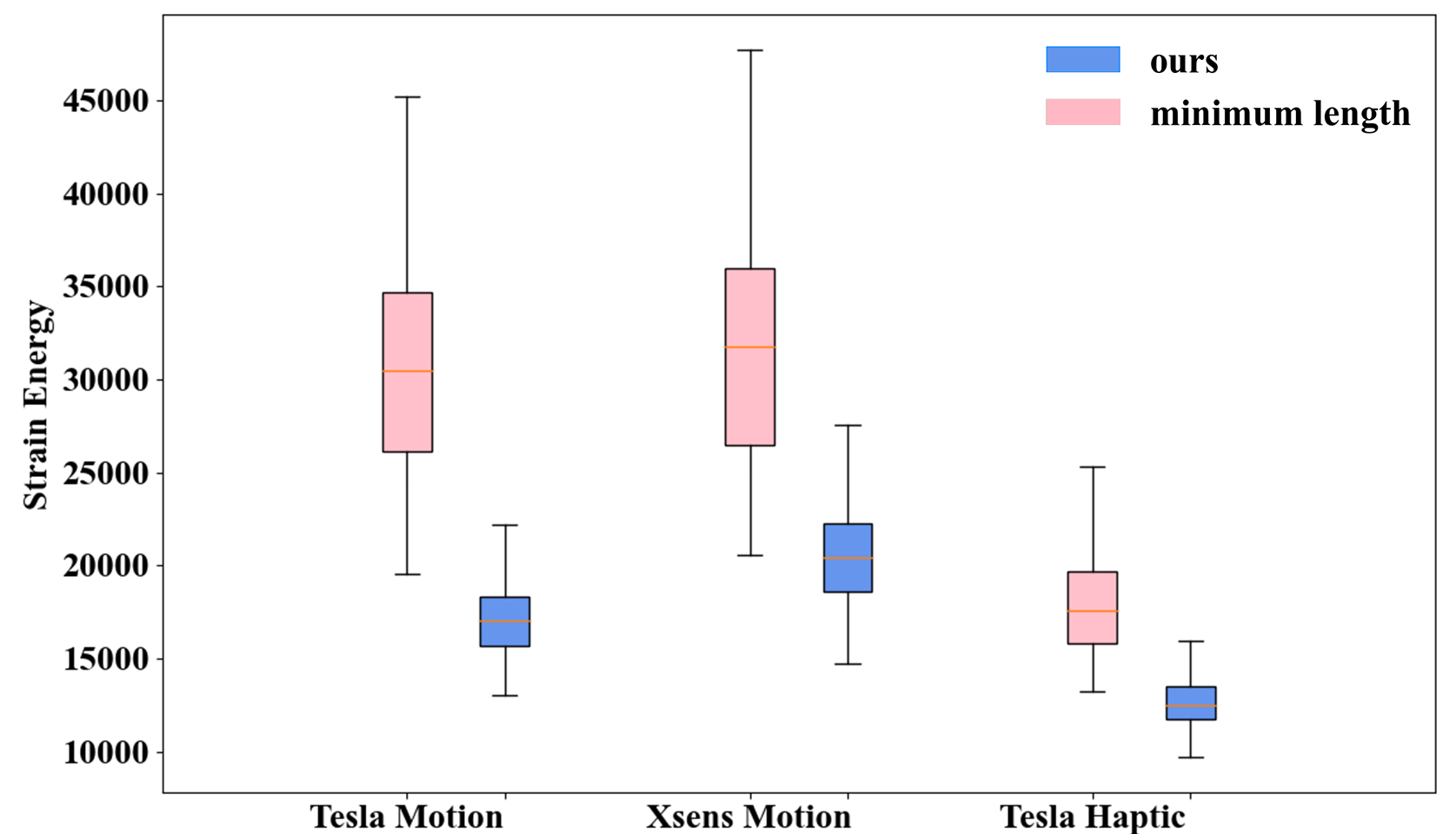}
         \caption{Strain Energy}
     \end{subfigure}
     \begin{subfigure}[b]{0.32\linewidth}
         \centering
         \includegraphics[width=\textwidth]{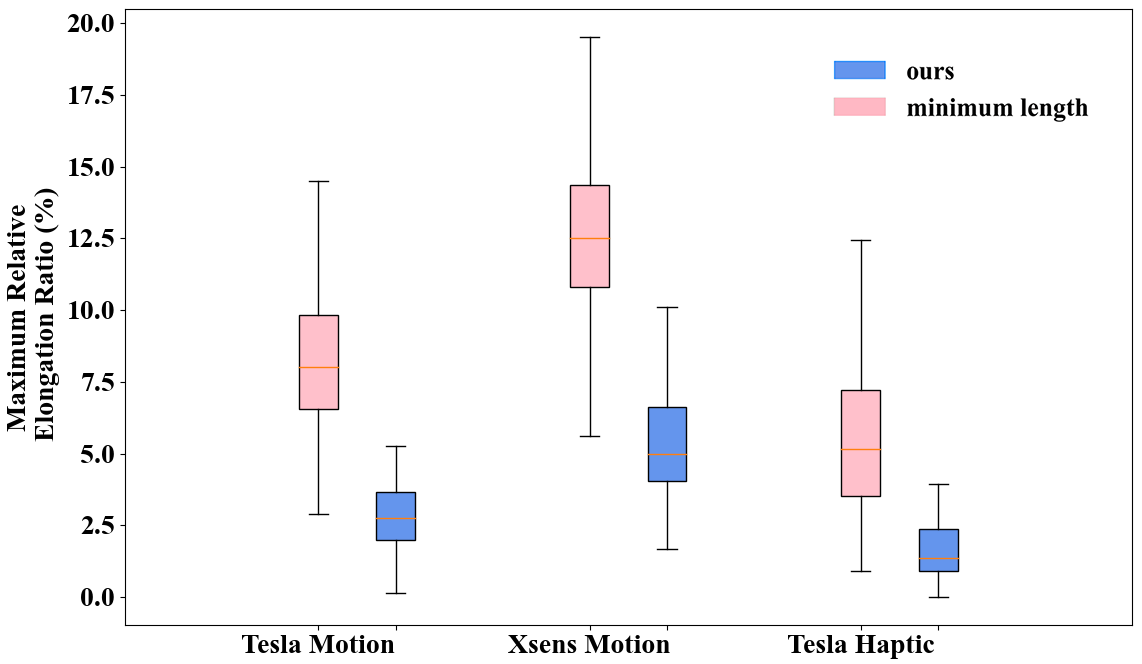}
         \caption{Maximum Wire Elongation}
     \end{subfigure}
     \begin{subfigure}[b]{0.32\linewidth}
         \centering
         \includegraphics[width=\textwidth]{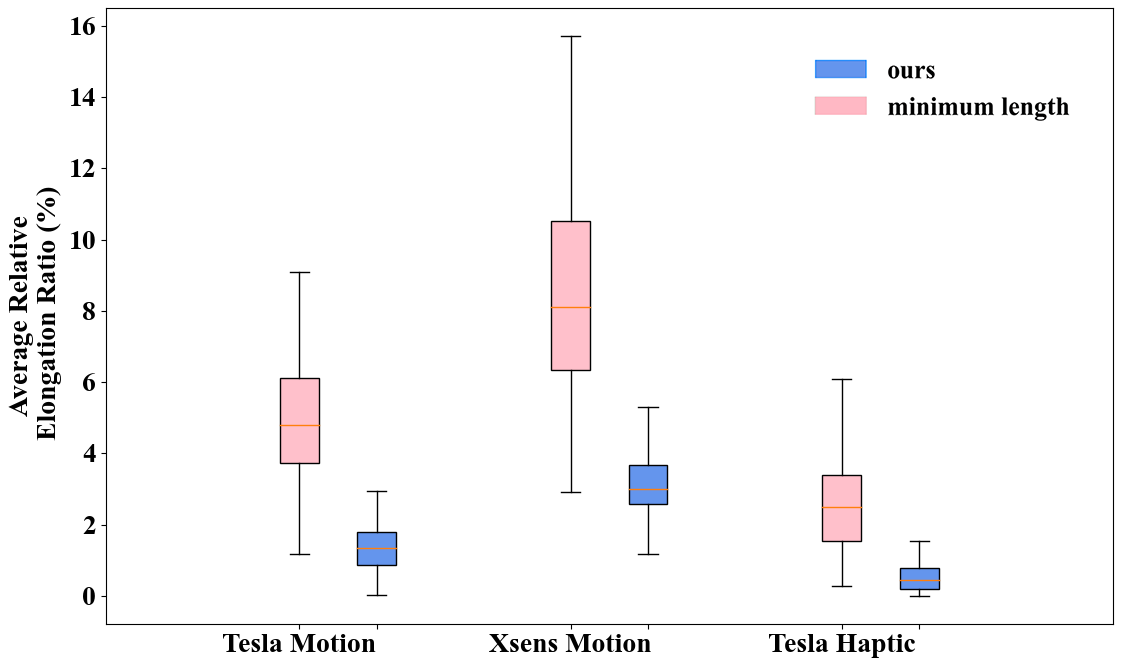}
         \caption{Average Wire Elongation}
     \end{subfigure}     
    \caption{Quantitative comparison between our method and the baseline under three terminal configurations.}
    \label{fig:comp}
\end{figure*}

As shown in Fig.~\ref{fig:comp}, i) our method proves effective in reducing wire strain on a Tesla suit, which features ten sensor locations covering every joint in the body. 
On average, our layout achieves a remarkable 43\% reduction in energy sum, coupled with reductions of over 65\% in both maximum and average stretch ratios (ave. 66.2\% and 70.6\%). 
ii) For the Xsens motion capture suit, our layout achieves an average 35\% reduction in energy sum and over 55\% reductions in maximum and average stretch ratios (ave. 55.7\% and 61.9\%).
iii) As a more challenging case, the Tesla Haptic Feedback configuration deliberately avoids stretching areas to some extent by carefully positioning the sensors. 
Despite this configuration, our layout outperforms the minimum length baseline, resulting in a smaller stretch. 
Both the maximum and average stretching are reduced by over 65\% (ave. 66.7\% and 77.5\%).

\subsection{User Experiment, Evaluation \& Feedback}
We conducted two user experiments to evaluate both the design process and results: one with expert designers (simulation) and another with clothing wearers (real-world experiment using capacitive sensors). 
The experiment received ethical approval from the authors' institution's ethics committee. All participants were provided with information about the experiment's purpose and procedure, and they signed an agreement prior to participation.

\subsubsection{Design Experiment of Experts}

\paragraph{Participants}
Two experts in the field of smart clothing were invited to design the wiring layout.
One expert is an industry designer with 23 years of experience, while the other expert is an associate professor and dean in the department of clothing and fashion design at a university.
Both experts have more than 5 years of experience working with smart clothing, in commercial products and research prototypes respectively.

\paragraph{Procedure}
We used the two commercial sensor configurations (Xsens, Tesla Suit Motion Capture) mentioned in the previous section as a reference. 
The experts were tasked with designing a wiring layout that aimed to minimize stretching and redundancy for the types of actions in our motion dataset, including dancing, boxing, golfing, football, and other sports (see Supplementary Sec.~4). 
Each design for the two reference configurations took approximately 3 days.
We evaluated the performance of both the expert layout and our layout, and compared the differences in the layout through feedback obtained from user experiments.
Following the experiment, the experts participated in a semi-structured interview to gather their subjective feedback.

\paragraph{Results \& Findings}
As Table~\ref{tab:user} shows, in the simulation, our layout continues to outperform the expert layout in terms of reducing deformation energy and the maximum wire elongation rate. Moreover, our wiring algorithm is significantly faster than expert designs.
\begin{table}[h]
  \caption{Comparison with expert design (simulation). Min-Len: our minimum-length baseline.}
  \label{tab:user}
  \begin{tabular}{c|P{1.8cm}P{2.2cm}P{2.5cm}}
    \toprule
    &Deformation Energy $\downarrow$ & Max Elongation Ratio (\%) $\downarrow$& Average Elongation Ratio (\%) $\downarrow$  \\
    \midrule
    Expert  & 25,404.92  & 6.39 & 3.92\\
    {\bf Ours}  & \textbf{20,753.09}& \textbf{5.63} & \textbf{3.22} \\
    \midrule
    Min-Len & 31,986.91 & 12.72 & 8.45 \\
  \bottomrule
\end{tabular}
\end{table}
During the interview, experts acknowledge that our layout effectively avoids significant areas of stretching and aligns with their design expertise. 
It is also worth noting that based on our pattern diagram, manufacturers can swiftly capture the necessary production information within the clothing pattern.

\subsubsection{Try-on Experiment of Participants}
\paragraph{Participants}
We recruited a total of 10 participants for this experiment, comprising undergraduate and graduate students from our institution. The participants' ages ranged from 21 to 35 years. Their height varied from 168 cm to 178 cm, and their weight ranged from 61 kg to 72 kg. All participants exhibited normal motor capability and reported no motor injuries within the past six months.

\paragraph{Procedure}
We evaluated three layouts: minimum length (baseline), ours, and the expert layouts. Tight suits with these layouts were fabricated in the standard size (175cm, 75kg). The participants were unaware of the layout type during the evaluation. The order of try-ons was counterbalanced. We conducted tests involving routine movements and free play movements. Participants rated the resistance and comfort for each layout using a 5-point Likert scale questionnaire. After the experiment, participants were interviewed to gather their subjective feedback.

\paragraph{Results \& Findings}

\begin{figure}[h]
    \includegraphics[width=0.9\linewidth]{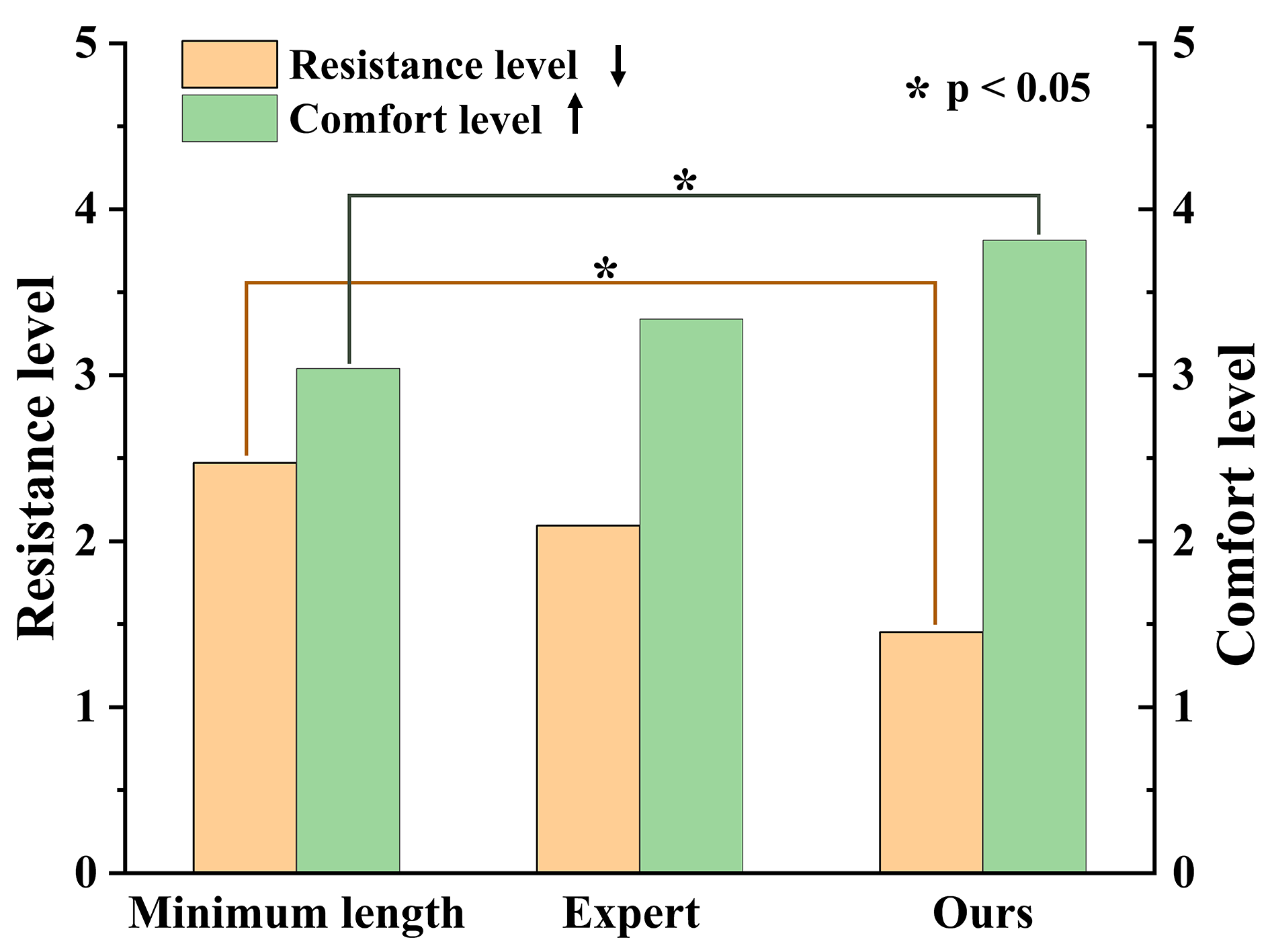}
    \caption{Average user scoring. Note that for the criteria of the resistance level, a lower value is better. The opposite holds for the comfort level.}
    \label{fig:userstudy}
\end{figure}

The experimental results indicate that our layout performed slightly better than the expert layout and significantly better than the minimum length baseline in terms of user scoring (Fig.~\ref{fig:userstudy}). One-way ANOVA analysis revealed a significant difference between our method and the minimum length baseline in terms of both resistance (p-value: 0.047, F-score: 4.559) and comfort (p-value: 0.026, F-score: 5.860) levels. Interestingly, there was no significant difference between our layout and the expert layout, although our layout exhibited a moderate numerical advantage over the expert layout.

\subsection{Discussions}

\begin{figure}[h]
    \includegraphics[width=0.95\linewidth]{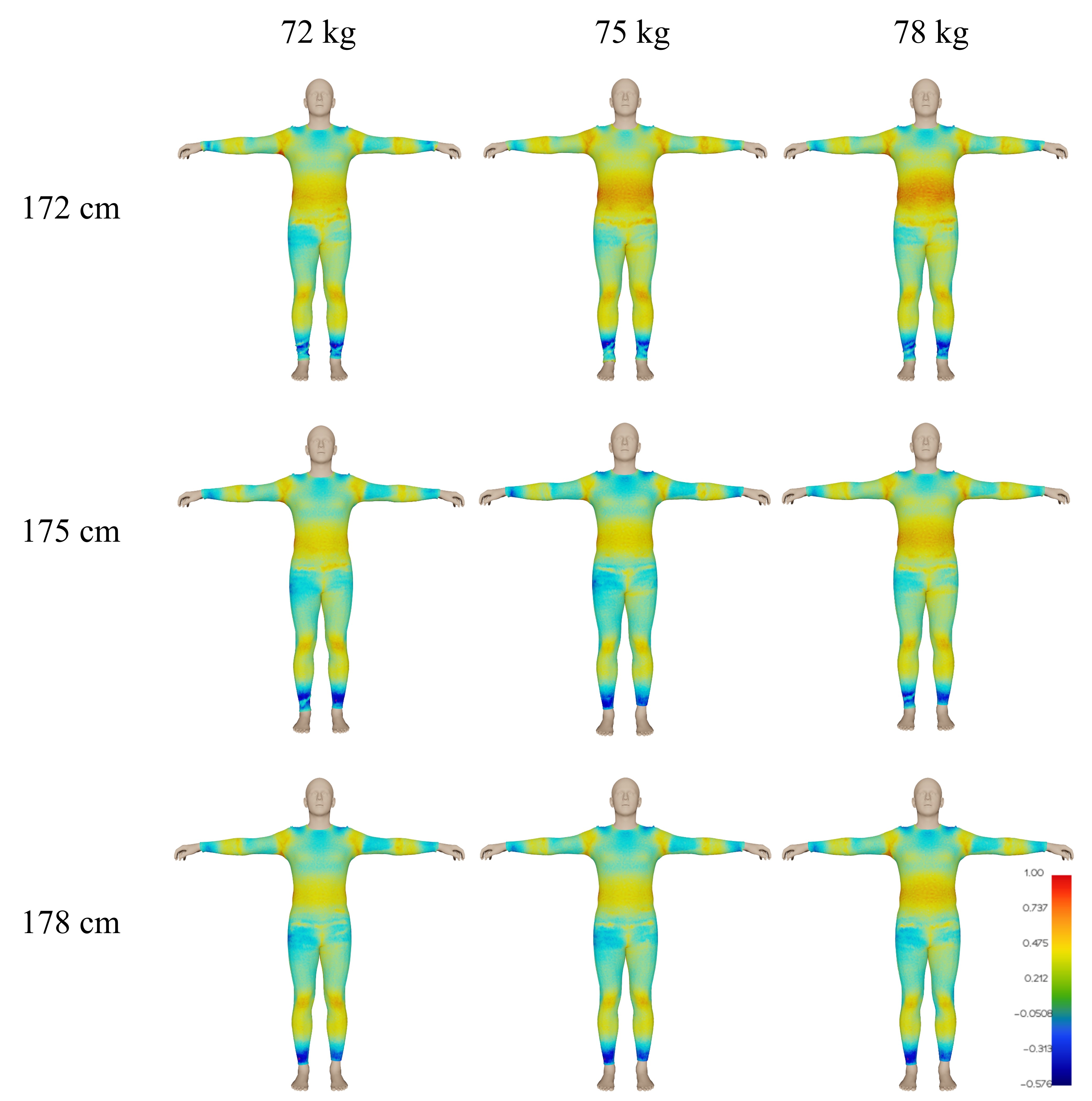}
    \caption{Strain deformation energy for different body weights and heights, given the same motion dataset $M$.
    The strain deformation demonstrates similar distribution, however, for a shorter body but with the same/a larger weight (172 cm/75 kg or 172 cm/78 kg), the magnitude on the belly is considerably higher than others. 
    This is reasonable, as these two configurations indicate a more obese body shape.
    }
    \label{fig:bodyweight}
\end{figure}

\begin{figure}[h]
    \includegraphics[width=0.92\linewidth]{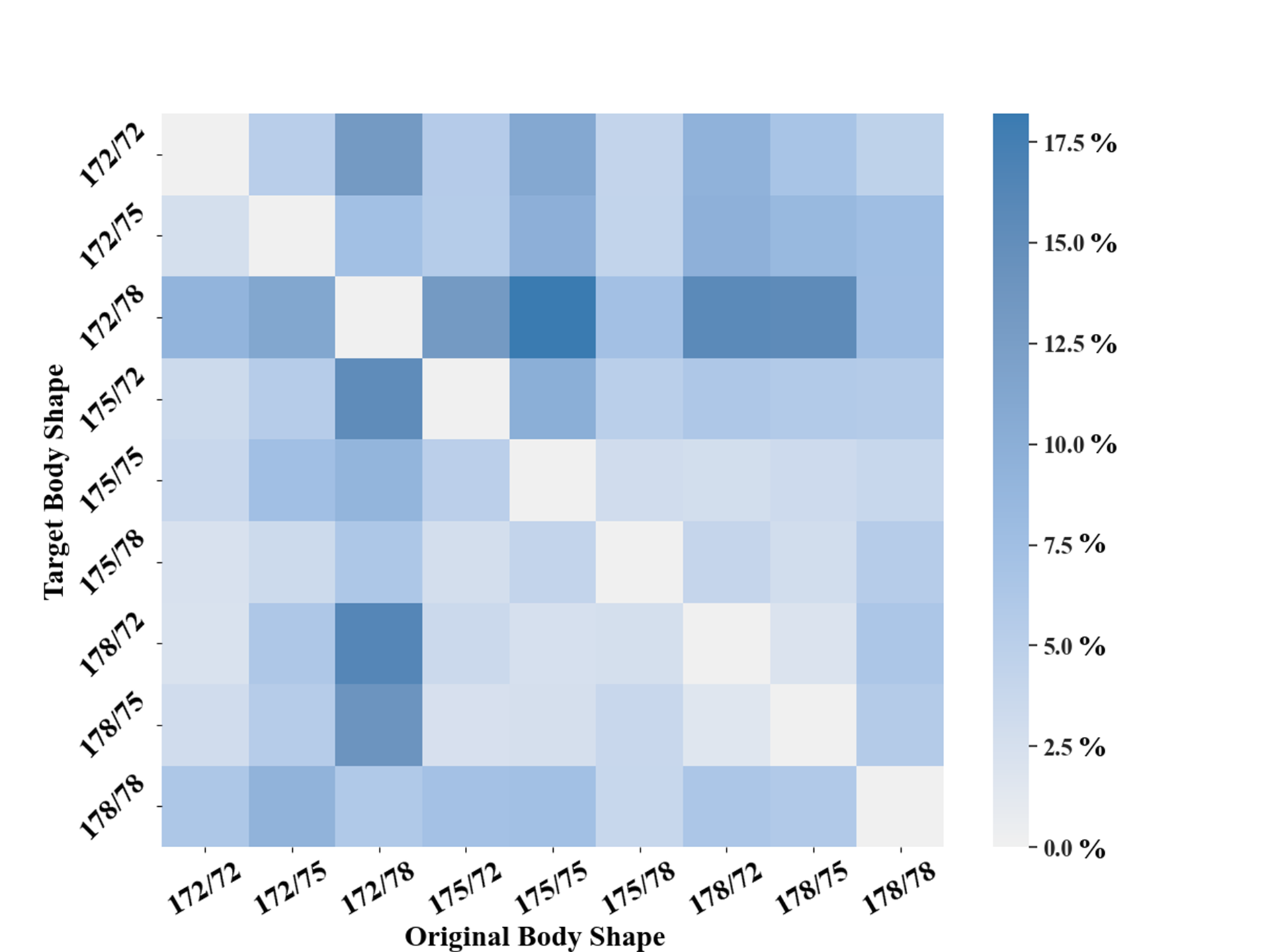}
    \caption{The ratio of energy difference between the layouts of different body shapes, and that of the corresponding body shapes.
    The largest difference ratio is 17.8\%, which indicates an acceptable generalization over an appropriate range of body variations (note that the difference between our design and the expert design is $\sim$18\%).
    }
    \label{fig:bodyweight-matrix}
\end{figure}

\begin{figure}[h]
    \includegraphics[width=0.98\linewidth]{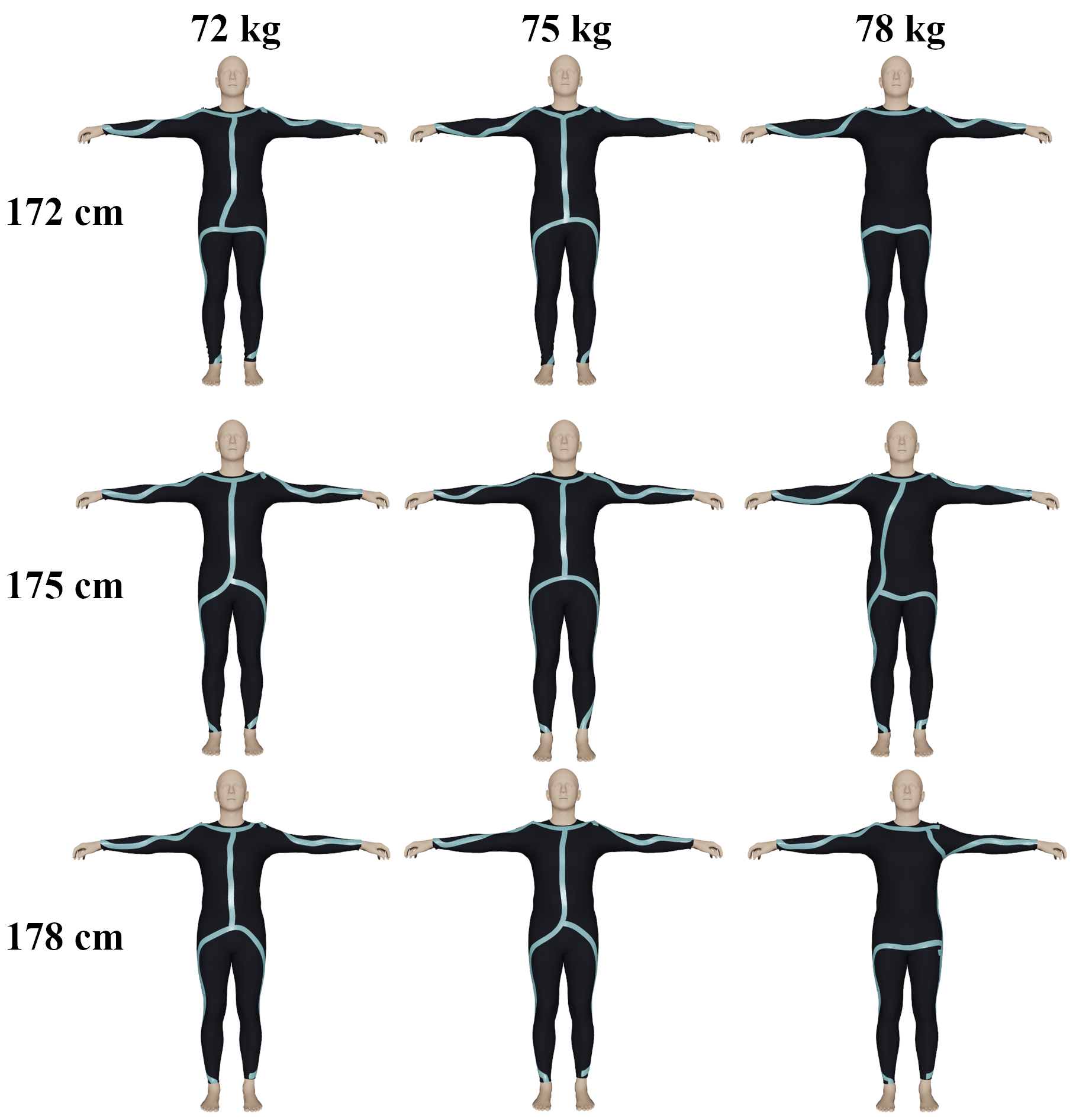}
    \caption{Wire layout for different body weights and heights.
    The layout design follows the pattern of strain deformation energy and presents a high similarity for most body shapes.
    An apparent difference is the belly region for body shapes with a larger weight.
    It is worth pointing out that although the layout demonstrates a certain variation, the ratio of deformation energy difference is relatively small (see Fig.~\ref{fig:bodyweight-matrix}). 
    }
    \label{fig:bodyweight-layout}
\end{figure}

\paragraph{Impact of Body Shape}
We conducted an analysis of the optimal layouts for similar but different body shapes performing the same action. Using the standard SMPL model, we generated eight body types (175$\pm$3 cm, 75$\pm$3 kg) that closely resemble the standard body type (175 cm, 75 kg) (Fig.~\ref{fig:bodyweight}). The strain energy distribution exhibited similarities across these body types, with minor variations in magnitude. After generating the optimal layout for each body shape, we cross-validated the layouts on all body types and calculated the deformation energy values. The results depicted in Fig.~\ref{fig:bodyweight-matrix} show that the differences between the calculated layouts for different body types were below 20\% in terms of their deformation energy. Furthermore, the experimental results indicated that the optimal layouts for body weights below 75 kg exhibited significant similarity, while the layout for 78 kg body weight showed slight differences compared to other layouts (Fig.~\ref{fig:bodyweight-layout}).

\begin{figure}[h]
    \includegraphics[width=\linewidth]{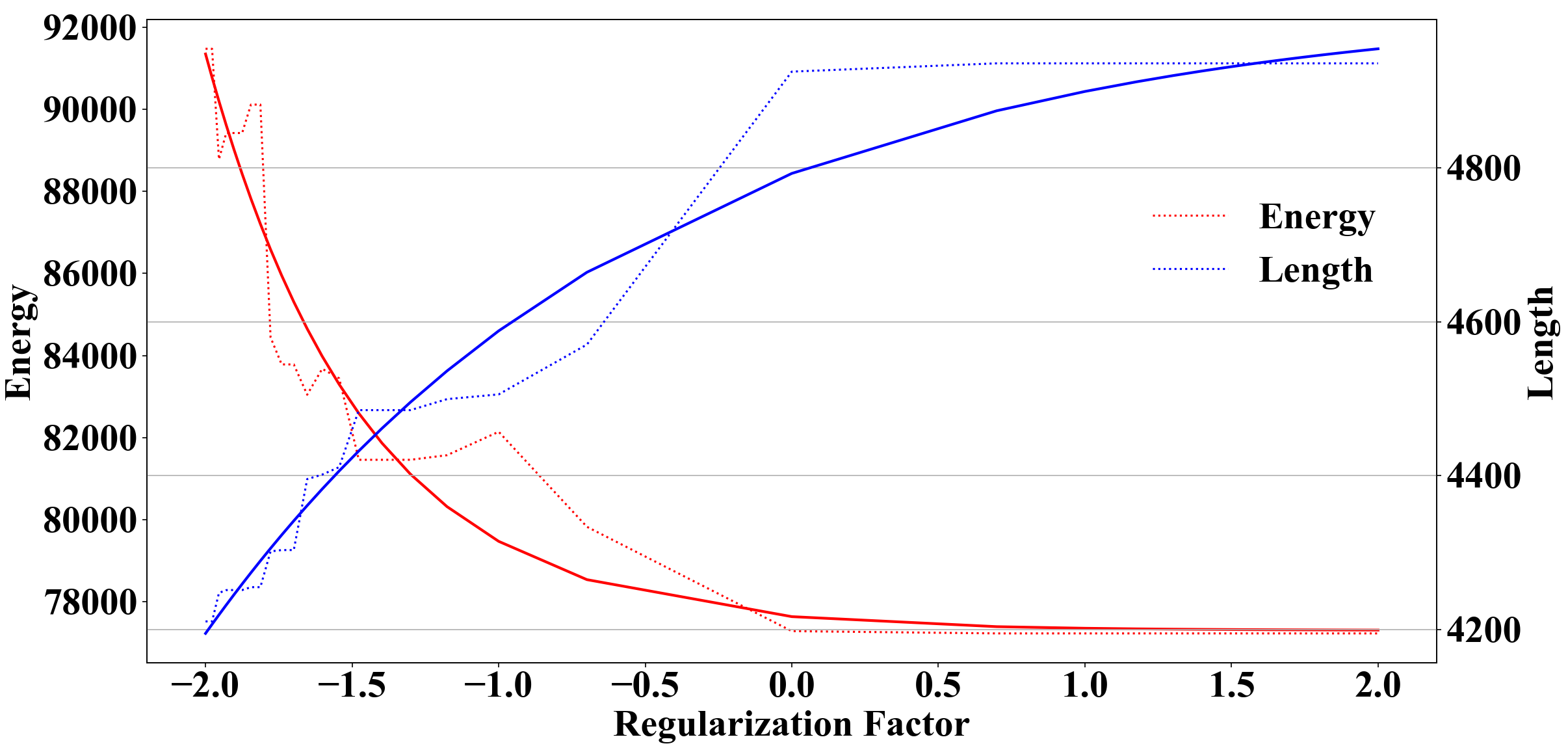}
    \caption{Total wire length and strain energy under various regularization conditions.
    The horizontal axis is the value of $-log_{10}(\eta)$.
    As $\eta$ increases, the deformation energy increases while the total length decreases.
    Equivalently speaking, a shorter wire path is most likely to cause a larger value of strain deformation and a higher level of motion resistance.
    An optimal balance arises when two curves cross in the above figure.
    The unit for length is millimeter.
    }
    \label{fig:regular}
\end{figure}

\paragraph{Impact of Regularization Factor}
The regularization term ${\eta}$ in Eq.~\ref{edgeweight} affects the automatic layout result. 
As depicted in Fig.~\ref{fig:regular}, increasing the value of ${\eta}$ leads to a reduction in the total wire length but an increase in strain energy. 
The layout reaches convergence when $\frac{1}{\eta} > 1$.

\paragraph{Impact of Mesh Resolution \& Terminal Numbers}
The precision and time cost of our wiring solution are influenced by two key factors: the number of terminals and the mesh resolution. As discussed in Sec.~\ref{sec:layout}, the discretization of the surface can result in a slight loss of precision and affect the time required to solve the STP. In this experiment, we use the centers of the triangles to subdivide the mesh at different levels. While the mesh resolution has a minimal impact on the precision ($<1\mille$) when solving the STP problem with the same number of terminals ($N_p$), it significantly affects the time performance, with differences of an order of magnitude (Table~\ref{tab:time_value}). With the current mesh resolution (e.g., $N_V$=23,515 in our case), the time performance remains at an interactive rate, regardless of the variation in the number of terminals $N_p$.

\begin{table}[h]
  \caption{Time (unit: second) and sum of edge weight for different number of terminals $N_p$ and mesh vertexes $N_V$.}
  \label{tab:time_value}
  \centering
  \resizebox{\columnwidth}{!}{%
  \begin{tabular}{P{0.2cm}P{1.1cm}|P{2.0cm}P{2.0cm}P{2.12cm}}
    \toprule
    &\multirow{2}{1.2cm}{\hfil Time (Weight)}&\multicolumn{3}{c}{$N_p$}\\
     && 10 & 15 & 36  \\
    \midrule
       
     \multirow{3}{0.2cm}{$N_V$}&23,515& 0.43 (4,360.48
)&0.47 (4,067.24
)&1.80 (5,435.95
)
\\
    &164,251& 4.152 (4,360.46
)&5.20 (4,067.23
)&20.34 (5,435.92
)
\\
    &727,195&23.43 (4,360.41
)&25.46 (4,067.18
)	&109.05 (5,435.86
)\\    
  \bottomrule
\end{tabular}
}
\end{table}

\paragraph{Limitations \& Future Work}
Computer-aided clothing design is an emerging topic~\cite{muthukumarana2021clothtiles,albaugh2019digital,kaspar2019knitting,mccann2016compiler,narayanan2018automatic,narayanan2019visual,zhang2019computational,wu2019knittable}, and our work extends to a more focused category: smart clothing. 
One potential area to improve our method is the discretization in our current solution, which involves solving the STP on a mesh as an approximation to that on a surface. As a result, the wiring layout is represented by a set of mesh edges rather than a smooth curve. 
A post-processing step can be applied to slightly refine the results.
In our experiments, the difference in deformation energy before and after post-processing is 4.78$\pm$2.36\%. Future work could aim to develop an integrated solution that combines wire layout optimization with curve generation, potentially achieving more accurate and refined results.
Also, our algorithm decouples strain computation from wiring layout optimization. Although decoupling these steps provides computational advantages, it fails to account for how wiring layouts influence strain energy in the fabric. Future work could investigate coupling strain computations with wiring optimization to better capture wiring effects on strain and user comfort.
Finally, our work focused on wire stretching, while other factors that affect wearing comfort or damage the wire (e.g., wire bending) will also be interesting for future work.

\section{Conclusion}
Our work addresses the challenge of automatic wiring on tight suits to minimize motion resistance for users. We formulate the wiring as a Steiner Tree Problem on the clothing mesh and propose an energy function to evaluate wire stretching. We fabricate the clothing based on the generated designs and conduct real-world experiments, collecting qualitative subject feedback and quantitative sensor measurements. The results demonstrate that our designs are effective in reducing wire strain energy by an average of 77\% across 248 different actions.

\section*{Acknowledgement}
This work is supported by National Natural Science Foundation of China (62072383), the Fundamental Research Funds for the Central Universities (20720210044). This work is partially supported by Royal Society (IEC\textbackslash NSFC\textbackslash211022). Xiaohu Guo was partially supported by National Science Foundation (OAC-2007661).

\bibliographystyle{ACM-Reference-Format}

\newpage

\clearpage

\afterpage{\FloatBarrier}

\afterpage{\FloatBarrier}

\end{document}